# Synchronous post-acceleration of laser-driven protons in helical coil targets by controlling the current dispersion

Zhipeng Liu[1], Zhusong Mei[1], Defeng Kong[1], Zhuo Pan[1], Shirui Xu[1], Ying Gao[1], Yinren Shou[1], Pengjie Wang[1], Zhengxuan Cao[1], Yulan Liang[1], Ziyang Peng[1], Jiarui Zhao[1], Shiyou Chen[1], Tan Song[1], Xun Chen[1], Tianqi Xu[1], Xueqing Yan[1,2,3], and Wenjun Ma[1,2,3]

[1] *State Key Laboratory of Nuclear Physics and Technology, and Key Laboratory of HEDP of the Ministry of Education, CAPT, Peking University, Beijing 100871, China*

[2] *Beijing Laser Acceleration Innovation Center, Huairou, Beijing 101400, China*

[3] *Institute of Guangdong Laser Plasma Technology, Baiyun, Guangzhou 510540, China*

**Abstract** Post-acceleration of protons in helical coil targets driven by intense, ultrashort laser pulses can enhance the ion energy by utilizing the transient current originating from the self-discharging of the targets. The acceleration length of the protons can exceed a few millimeters, and the accelerating gradient is in the order of GeV/m. How to ensure the synchronization of the accelerating electric field with the protons is a crucial problem for an efficient post-acceleration. In this paper, we study how the electric field mismatch induced by the current dispersion affects the synchronous acceleration of the protons. We propose a scheme using a two-stage helical coil to control the current dispersion. With optimized parameters, the energy gain of protons is enhanced by 4 times. And it is expected that the proton energy would reach 45 MeV using a hundreds-terawatt laser, or over 100 MeV using a petawatt laser, by controlling the current dispersion.

___________

Correspondence to: Wenjun Ma, State Key Laboratory of Nuclear Physics and Technology, Center for Applied Physics and Technology, School of Physics, Peking University, Beijing 100871, China. Email: wenjun.ma@pku.edu.cn

# 1. Introduction

Ion acceleration driven by intense and ultrashort laser pulses[1, 2] has attracted more and more attention as it can produce MeV-GeV ions in micrometer scale. The unique properties of the laser-driven ions, such as ultra-high peak current[3, 4] and small source size[5], which are in high demand for applications such as radiography[6], FLASH radiotherapy[7, 8], material science[9] and nuclear fusion[10]. Among the acceleration mechanisms proposed so far, the most rigorously studied mechanism is the target normal sheath acceleration (TNSA) [11-13]. When an ultra-intense laser pulse irradiates a thin solid target, energetic electrons are generated on its front surface. Those hot electrons then penetrate the target and create a substantial charge-separation electric field on the target's rear surface. Ions can be accelerated in this electric field to high energy in a few micrometers. It has been identified as a robust method that can stably generate proton beams with maximum energy up to several tens of MeV[14]. However, the theoretical models[15] and experimental results[16, 17] prompt the scaling law between TNSA accelerated ion beams energy and laser intensity as $E_{ions} \propto I_{laser}^{1/2}$, indicating that it is highly difficult, if possible, to obtain hundreds-of-MeV protons at currently available laser intensity[18-20].

To overcome the predicament of TNSA acceleration, Kar *et al.*[21] proposed and realized a scheme for simultaneous post-acceleration, energy selection, and collimation of ions by attaching a sub-millimeter-diameter helical coil (HC) normal to the rear side of the metallic target foil. It is a further utilization of the escaped electrons from the laser-ion acceleration without needing an extra laser pulse. In a typical TNSA, the escaped electrons have a picosecond scale duration and charge of a few to hundreds of nC[22]. As the result of the escaped electrons, a transient current is born in the HC in the form of a surface wave, which propagates along the coil at a speed close to *c*, the speed of light in the vacuum[23, 24]. This current has an ultra-high density up to the order of $\mu Cm^{-1}$, which can build a transient electric field with the magnitude of GV/m in the center of the HC[21] to post-accelerate the TNSA proton beam. In the demonstration experiment of the HC post-acceleration (HCPA), where a 200 TW laser system was employed, the maximum proton energy was enhanced by 35% (2.7 MeV) using a 7-mm-long HC. The later work by Ahmed *et al.*[25] (2017) highlighted the function of the HCPA in beam focusing and energy spectrum modulation. They obtained a highly collimated (<1° divergence cone) and narrow-band (~10% energy spread) proton beam at ~10 MeV. The energy gain can be more significant with increased laser energy and intensity. Simulations based on a test-particle approach predict that a single HC can boost the proton energy by 30 MeV with a PW laser[21]. However, the experimental results showed that the energy gain was not as high as expected: merely 12 MeV enhancement was observed instead of the 30 MeV[26].

Prolonging the acceleration distance by maintaining the synchronization of the HCPA field and protons would be a practical way to increase the energy gain in experiments. Kar *et al.*[21] conceived a two-beam laser-triggered HC's current, allowing the protons to travel through the HC twice in sequence, and the maximum energy in the simulation exceeded 100 MeV. Moreover, it was proposed that the acceleration can be adapted to the increased speed of the ions by adjusting, continuously or stepwise, the pitch and radius of the HC[26]. However, the experimental energy gain was significantly lower than the theoretical prediction, and increasing the length of the HC did not give a further increase in energy gain.

In the above works, the current propagating in an HC was assumed to be dispersionless. As a matter of fact, the picosecond transient current has a frequency spectrum as broad as tens of GHz[27-30]. When such a broadband current propagates along an HC, the dispersion affected by the coupled inductances and capacities is nonnegligible. Bardon *et al.*[31] have found in the simulation that the velocity dispersion of the current leads to progressive modification of the HCPA, which might be one of the most important reasons for the termination of the acceleration in a long HC. Since the previous experiments and simulations did not consider the current dispersion, deep investigation of its effect for a long-time synchronization between the post-acceleration field and the protons is highly demanded.

In this paper, we perform a systematic study on the HCPA considering the dispersion effect by employing self-consistent electromagnetic field and beam dynamics simulations. We first illustrate the relationship between the dispersed current and the electric field and its impact on the proton beam dynamics, which provides important insight into the HCPA. It is found that the sudden phase reversal of the electric field induced by the current dispersion is the primary reason for the termination of the HCPA. Based on our understanding, we propose a two-stage HC structure to compensate for the phase change. Under controlled synchronization of the acceleration field with the protons, the gain of the HCPA is enhanced by four times.

The article is structured as follows: Sec. 2 describes the setup and method of the large-scale self-consistent simulations for the current and the electromagnetic field (EMF). In Sec. 3, we present the dispersion of the current in a straight wire and an HC and explain it with a circuit transmission model. In Sec. 4, we analyze the evolution of the electric field and the post-acceleration of protons in a single-stage HC, and demonstrate that how the current dispersion leads to asynchrony of the electric field and proton beam. In Sec. 5, we propose a dispersion-controlled two-stage HC scheme to enhance the energy gain. In Sec. 6 we discuss the scheme using multi-stage HC (more than two), and find that additional stages are not beneficial for further energy enhancement. Sec. 7 summarizes our results.

## 2. Simulation method

Several numerical simulation methods of the HCPA have been reported previously. Jiang *et al.*[32] executed the particle-in-cell (PIC) simulations about the laser's interaction with an ionized tens micrometer solenoid target. However, these explicit PIC simulations would be unacceptable for such a large HC target (a few centimeters) because of computational limits. Kar *et al.*[21, 25, 26, 33] used the particle-tracing code to investigate the proton dynamics, while ignoring the nonideality of the helix system. Bardon *et al.*[31] used finite-difference time-domain (FDTD) codes[34, 35] to simulate the current propagation through HC, which is a suitable approach to simulate the EMF in full-scale (nanosecond and centimeter scale). However, their simulations lacked particle dynamics analysis to reveal the effect of velocity dispersion on post-acceleration.

Here, we employ the CST particle studio Suite[36], a method that combines the FDTD and the PIC codes to simulate the EMF and beam dynamics in HCPA. The PIC simulations initialize the velocity and position of the charged particles, then the Maxwell's equations are solved by using the FDTD methods to obtain the EMF efficiently. The EMF and the particle dynamics are self-consistently described because all the terms in the Maxwell equations are retained in the equation scheme[37]. The generation of fast electrons and protons caused by the laser-plasma interaction (LPI) at the irradiation spot, which may require explicit PIC simulation, is simplified as particle source emission with energy distribution and total number subject to the interaction mechanism[23]. As a result, the discharged current will be excited spontaneously and propagate following Maxwell equations. The interaction between the emitted particles and the generated EMF is self-consistent, and both spatial-temporal profiles of EM fields and proton dynamics could be specified.

In order to let the simulation closer to the actual process, we perform full-scale simulation in time and space amounting to the nanosecond and centimeter scale. The configurations of simulations are shown in Figure 1(a). The foil target (Au, 10 μm thickness) is attached coaxially to the HC (aluminum, 100 μm wire diameter). All the neighboring structures touch each other and are connected to the pure copper holder at the end to form a path for the current. The components of the target, coil, and holder are meshed in high density with the cell size $\Delta x = \Delta y = \Delta z = 10$ μm . The whole simulation volume is 45 cm³. Electrons and protons are modeled with 10 million macroparticles. The energy distribution of the proton beam is a Maxwellian function with an effective temperature $T_p = 3.5$ MeV and cut-off energy $E_{p,cut-off} = 25$ MeV . The energy spectrum of escaped electrons is a Maxwellian function with $T_e = 2.5$ MeV and $E_{e,cut-off} = 25$ MeV , close to the reported works[38, 39]. The total charge of the escaped electrons and protons is 132 nC and 10 nC, respectively. As predicted by the model in references[40, 41], the protons' energy and electrons' charge could be achieved using a hundreds-terawatt femtosecond laser at an intensity of $8\times10^{20}$ Wcm$^{-2}$ . The $dN_{p,e}/dt$ of emitted protons and electrons are taken to be Gaussian function with an FWHM duration of 1.0 ps[42], which is negligible compared to the total simulation time. As a result of the escape of the electrons and ions, a self-discharged current is born in the wire. Figure 1(b) shows the simulated temporal profile of the current. The peak charge density is 32 μCm$^{-1}$ and the pulse duration is 14 ps.

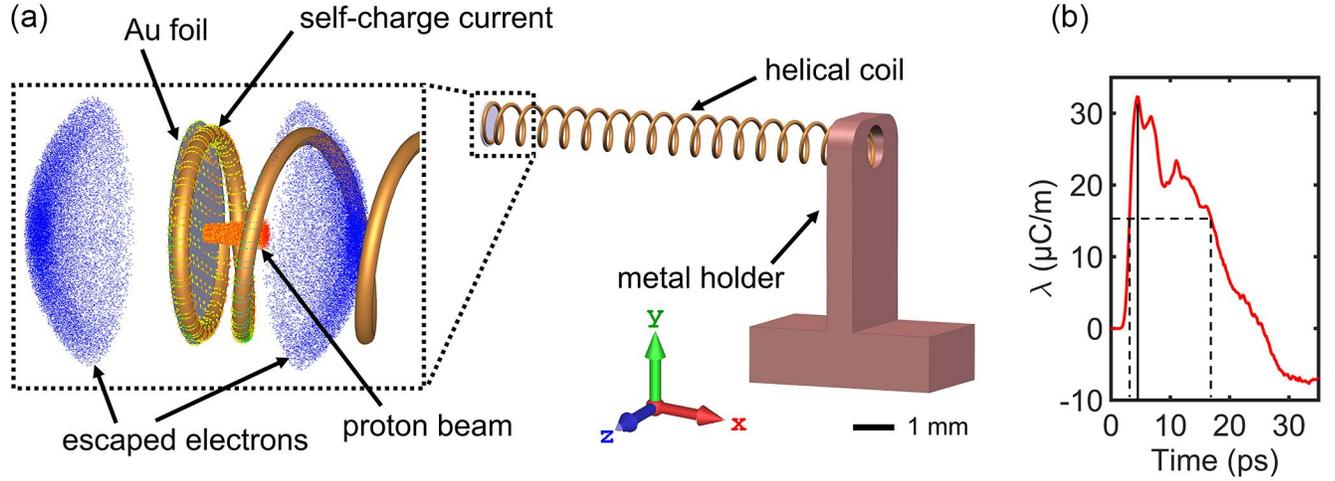

**Figure 1.** Simulation setup of (a) HC target configurations and (b) the self-discharged current generated by the emitted particles.

## 3. Dispersion of a transient current in a straight wire and an HC

Firstly, we simulate the propagation of a transient current on a straight Al wire. Figure 2(a) is the simulation result of the current as a function of time. The front of the current propagates along the wire at the speed of $c$. With its propagation in the $x$ direction, the tail of the current grows up due to dispersion. This can be numerically explained by the circuit transmission model[43]. The propagation of current can be described by the telegraph equation[44]:

$$\frac{\partial^2 J(x,t)}{\partial x^2} = LC \frac{\partial^2 J(x,t)}{\partial t^2} \tag{1}$$

where $J(x,t)$ is the current, $L$ and $C$ are the inductance and capacitance per unit length. Plugging in the propagating function of the current $J(x,t) = J_0 \exp(i(\omega t - kx))$ into the Eq. 1, one can obtain the dispersion relation for a transmission line:

$$k^2 = \omega^2 L_\omega C_\omega \tag{2}$$

For the Al wire, $L_\omega$ and $C_\omega$ weakly depend on ω. It can not be treated as an ideal transmission line mode when the ultrashort current signal propagates along a metal wire[45]. And the current dispersion should be considered. If the current propagates in an HC, the longitudinal component of the current can be written as $J_x(x,t) = J_{x0} \exp(i(\omega t - k_x x))$. $k_x$ is determined by the pitch and radius of the HC, as $k_x = \beta k$, where $\beta = p/2\pi a$ is called helix radio, $p$ and $a$ is the pitch and radius of the coils. Hence, the dispersion relation for an HC transmission system is:

$$k_x^2 = \omega^2 L_a C_a \quad (3)$$

where $L_a$ and $C_a$ are the equivalent inductance and capacitance of HC, which have been given by Kino and Paik in reference[31], as:

$$L_a = \frac{\mu}{2\pi} \frac{k_x^2}{\beta^2 \gamma^2} [I_1(\gamma a) K_1(\gamma a)]$$
$$L_a = 2\pi / [I_0(\gamma a) K_0(\gamma a)] \quad (4)$$

where $\gamma = (k_x^2 - \omega^2/c^2)^{1/2}$, $c = (\mu\varepsilon)^{-1/2}$, $I_n$ and $K_n$ is the modified Bessel functions of the first and second kind, respectively. Eq. 3 and Eq. 4 can be simplified as:

$$\frac{\omega}{k_x c} = \sqrt{\frac{\beta}{\beta^2 + D}}, \quad D = \frac{[I_1(\gamma a) K_1(\gamma a)]}{[I_0(\gamma a) K_0(\gamma a)]} \quad (5)$$

By solving Eq. 5 numerically, we can get the phase velocity $v_{ph} = \omega \lambda / 2\pi$ (see Figure 2(b)). The phase velocity will rise as the signal wavelength increases. In particular, if the wavelength is larger than the diameter of the HC, the dispersion is more severe, and the phase velocity would be higher than the speed of light in a vacuum. The strong dependency of $v_{ph}$ on $\lambda$ in the HC leads to an obvious dispersion-induced evolution of the current.

Figure 2(c) shows the spatial-temporal distribution of the current in a HC with pitch $p = 0.5$ mm and radius $a = 0.4$ mm. Unlike in the straight Al wire, the current dispersion in the HC is severe. The positive peak of the current becomes broader, and its amplitude decreases with time. Meanwhile, a negative current following the positive current emerges and grows with time (Figure 2 (d)). Similar phenomena alternately repeat with the propagation of the current in the HC, making the current waveform drastically varies. Technically, we can define the velocity of the main positive peak as the apparent longitudinal velocity of the current. According to the simulation results, it is about 1.6 $\beta c$ as shown by the green dashed line in Figure 2 (c). It should be noticed that this apparent velocity is greater than the expected velocity of $v_{ph} = \frac{p}{2\pi a}c = \beta c$ for a dispersionless current. In previous studies[25], the parameter of the HC was deliberately chosen so that $\beta c$ matches the protons' velocities. According to our simulation results, their strategy would inevitably lead to the desynchrony of the post-acceleration.

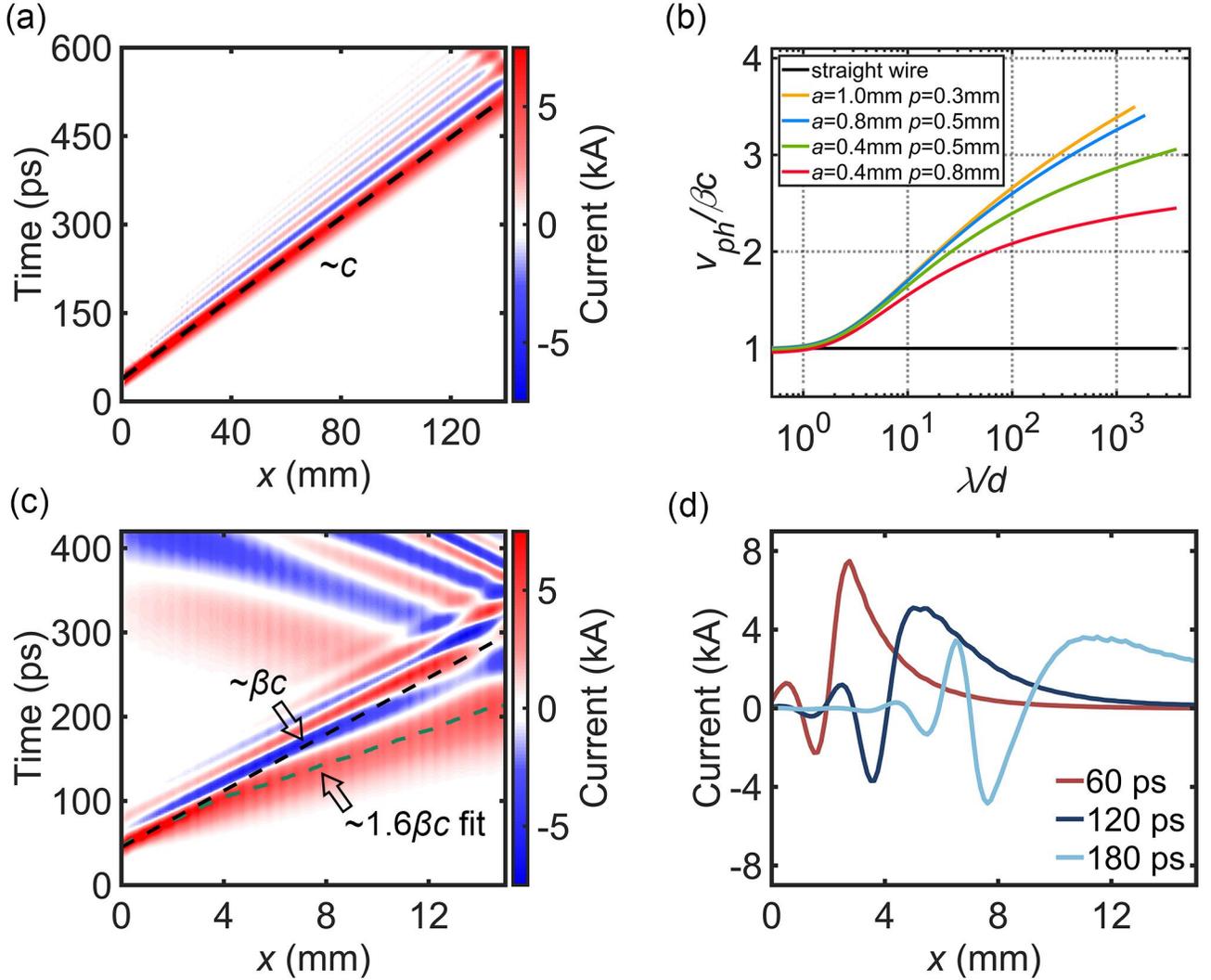

**Figure 2.** (a) Spatial-temporal distribution of the current on the straight aluminum wire, the black dashed line refers to the light speed *c*. (b) Dependence of the phase velocity and the ratio of wavelength to coil diameter ($\lambda/d$), with different radii and pitches in HC and straight wire. (c) Distribution of current in the HC, with the velocity mark of *βc* (black dashed line) and the fitting velocity 1.6 *βc* of the main positive peak (green dashed line). (d) Snapshot of current distributions in the HC at 60 ps, 120 ps and 180 ps.

## 4. Evolution of the electric field and the post-acceleration of protons in a single-stage HC

The traveling transient current builds a transient electric field around the wire. In the case of HC, the field near the center of the coil has a substantial longitudinal component accelerating the protons from the thin foil. In an ideal post-acceleration, the field witnessed by the protons should always be positive. The scheme of using varied pitch and radius of the HC can compensate for the phase sliding between the protons and the field to some extent. However, we found in the simulations that a sudden phase reversal of the field due to the current dispersion is more severe for a synchronous HCPA. Figure 3(a) shows the longitudinal electric field on the central axis of HC (labeled as $E_x$) obtained from the simulations. In the early time (60 ps), it is a dipole field resulting from the transient current. With the dispersion of the current, the field becomes multipole (120 ps, 180 ps). In Figure 3(b), if we look at the main peak of the positive field vanishes at $T = 80$ ps, and emerges 40 ps later but is delayed by $2\pi$ in phase. In other words, the phase of the leading extreme point is reversed. Such phase reversals occur two times in total in Figure 3(b) (shown by the three dashed lines on the positive $E_x$ field). Whenever it happens, protons that are accelerated a few ps ago would soon become decelerated. The fitting velocity of the positive electric field is 1.2 $\beta c$ in Figure 3(b). Due to the phase reversal, however, the protons cannot undergo a continuous acceleration even if their inject velocities are close to 1.2 $\beta c$.

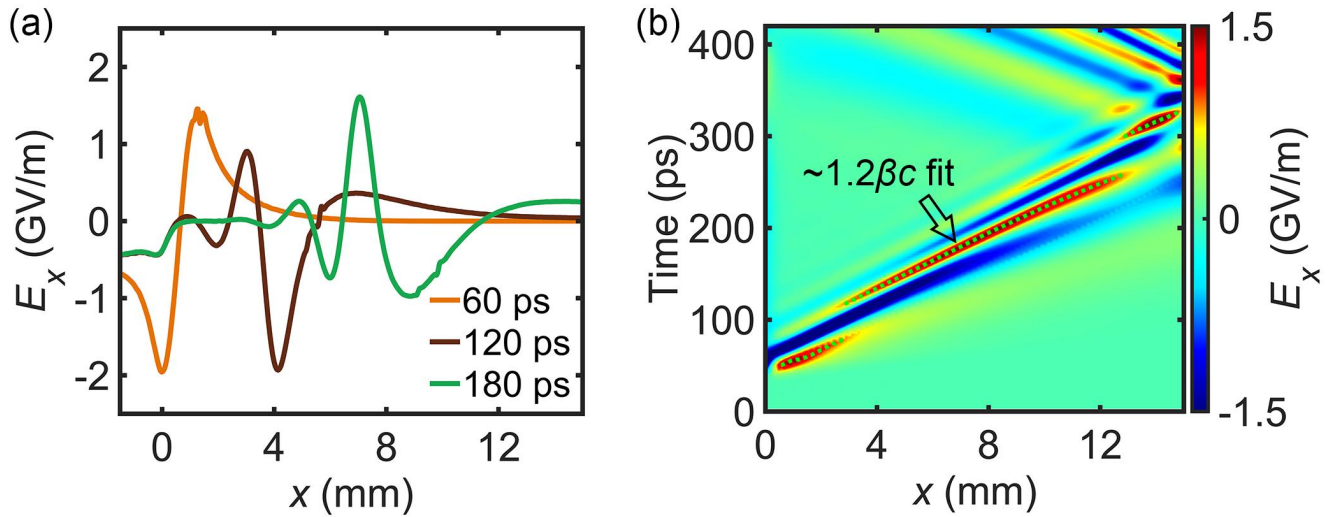

**Figure 3.** (a) Snapshot of the longitudinal electric field on the central axis of HC at 60 ps, 120 ps, and 180 ps. (b) Spatial-temporal distribution of the longitudinal electric field in an HC, with the mark of extreme points of positive fields (green dashed lines).

We numerically investigate the post-acceleration process of the protons in the evolving electric field in the HC. Figure 4(a) shows the snapshots of the protons' distribution in phase space and the longitudinal electric field at 60 ps, 240 ps, and 360 ps, respectively. Their initial maximum kinetic energy is 25 MeV. As we believe that the protons will be rapidly accelerated to match the speed of the positive electric field of 1.2 $\beta c$, corresponding to the 28 MeV protons. At $T = 60$ ps, the protons in the high-energy end located at the peak of the positive field undergo a most efficient acceleration. At $T = 240$ ps, the most energetic protons are however in the negative field. In general, most protons experience alternative fields and cannot continuously gain energy. At $T = 360$ ps, the cut-off energy of the protons is even lower than that at $T = 60$ ps. To investigate the energy evolution of the most energetic protons, we show their witnessed $E_x$ in their coordinate frame in Figure 4(b). The green circles represent their spatial positions in the above three snapshots. Firstly, they are post-accelerated to a maximum energy of 29.7 MeV at $x = 7$ mm in the positive field. Then they slide into the negative field and are decelerated to 24.4 MeV. With the reduction of their velocity, they are "caught up" by the positive field and accelerated again after $x = 20$ mm. Due to the alternative acceleration and deceleration, the eventual energy gain is relatively low. The highest energy gain for the most energetic protons, in our case, is 19%, which is comparable to the reported experimental results[21, 25, 26, 31].

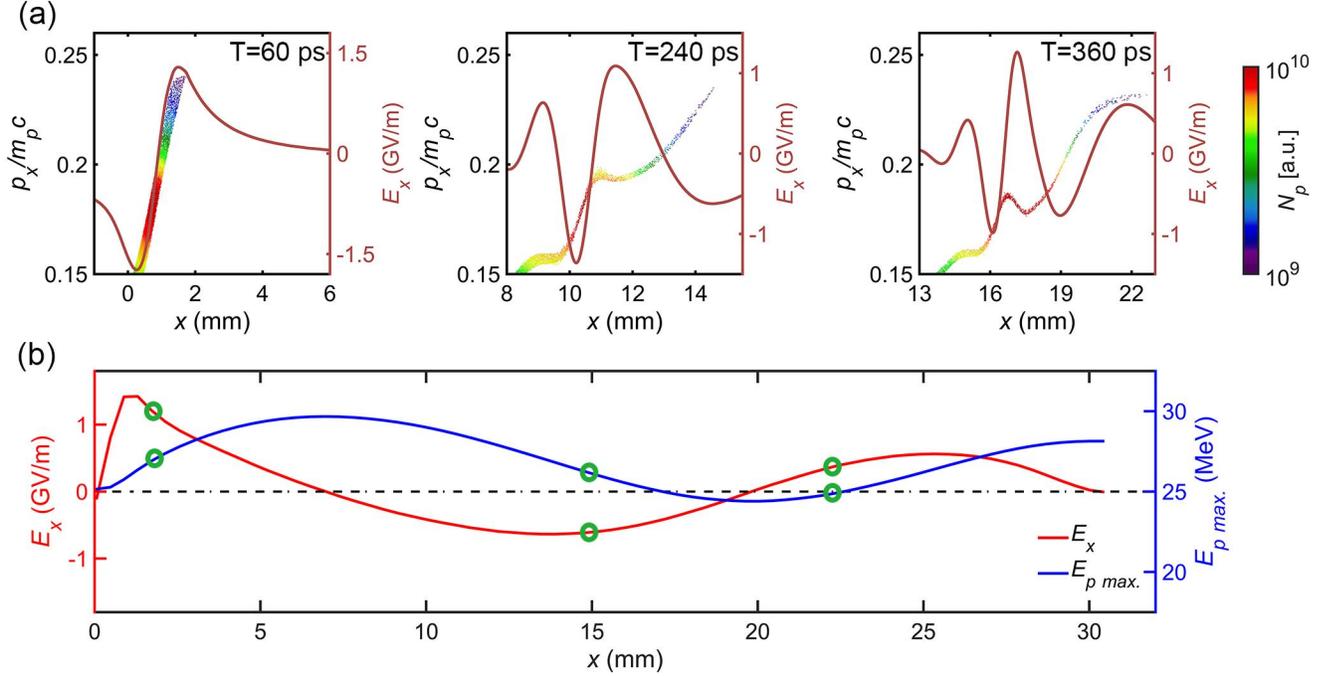

**Figure 4.** (a) Snapshots of the protons' distribution in phase space ($x$, $p_x$) and the longitudinal electric field at time of 60 ps, 240 ps and 360 ps, in single-stage HC. (b) Longitudinal electric field in the coordinate frame of the traveling highest-energy protons (red curve) and the evolution of the maximum energy (blue curve) in single-stage HC, the three groups of green circles mark the three statuses in (a).

We have simulated the energy gains of protons with different input energies from 20 MeV to 30 MeV in an HC, as shown in Figure 5(a). The protons in a wider energy range similarly undergo acceleration then deceleration. Figure 5(b) presents the $E_x$ distribution and the proton trajectories in it. It is shown that higher-energy protons will be "caught up" by the negative electric field more later, and will obtain higher energy gain, but it is still a faint enhancement.

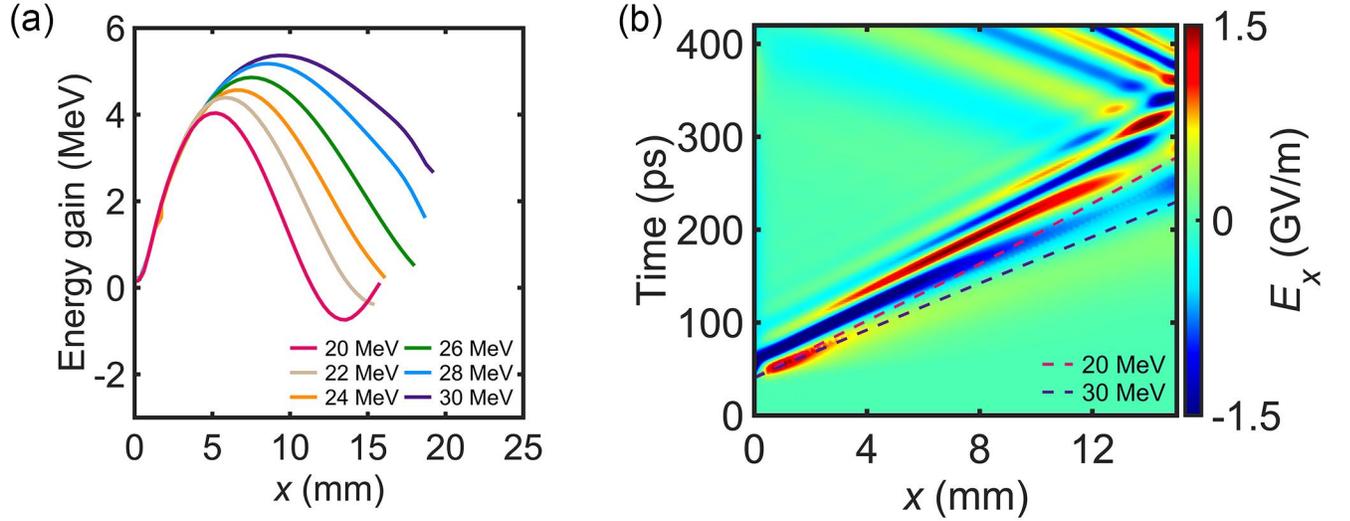

**Figure 5.** (a) Energy gains of traveling protons in HC with input energies. (b) The spatial-temporal distribution of $E_x$ and trajectories of protons with different input energies.

In summary, the deceleration stage resulting from the current dispersion drastically undermines the energy gain in the post-acceleration. The second positive electric field is not efficiently utilized. The existing methods, by continuously or stepwise adjusting the pitch and radius of the HC to extend the accelerated distance[26], cannot eliminate current dispersion and only obtain a short-time acceleration. New schemes that consider the current dispersion and can maintain a long-time synchronous acceleration are essential for a high-gain post-acceleration.

## 5. Enhanced post-acceleration in two-stage HCPA

To overcome the problems resulting from the current dispersion, we propose a scheme where a two-stage HC is employed, which can achieve a long-time synchronous acceleration for the protons. The geometry of the two-stage HC is shown in Figure 6(a). A straight wire is between the two coils as the "drift section". After the acceleration in the first HC, as shown above, the most energetic protons travel faster than the positive pole of the field and slide into the negative field. In the drift section, the current dispersion becomes milder, and the phase velocity of the current is higher than that of the protons. It is therefore possible to compensate for the phase sliding with proper parameters. After the drift section, the protons are in the positive field and start to be accelerated again in the second HC.

The drift section is introduced at the position of $x = 4$ mm, and the length of the drift section is chosen as 3.8 mm. The coil parameters are the same as that in a single-stage HC. The reason for the setting will be discussed later. Figure 6(b) and (c) show the spatial-temporal distribution of the current pulse and $E_x$ in the two-stage HC. In the first HC, the current undergoes a strong dispersion. But in the drift section, its waveform hardly changes as the dispersion is much weaker. Then it varies again after entering the second HC. In the drift section, the current propagates with the speed of $c$, much faster than the protons. The delayed positive peak thus is able to catch the protons again. As illustrated in Figure 6(c), the peaks of the positive field are marked by the green dashed lines. The positive electric fields in the first and second sections are seamlessly connected by a straight velocity line of 1.2 $\beta c$. Figure 6(d) shows the distributions of $E_x$ and the position of the protons with an energy of 25 MeV initially at different times. At 60 ps, the $E_x$ and the proton distribution are the same for the two cases. After the drift section, at 240 ps, the $E_x$ in the two-stage HC shifts forward relatively. The protons are "caught up" by the accelerating phase again.

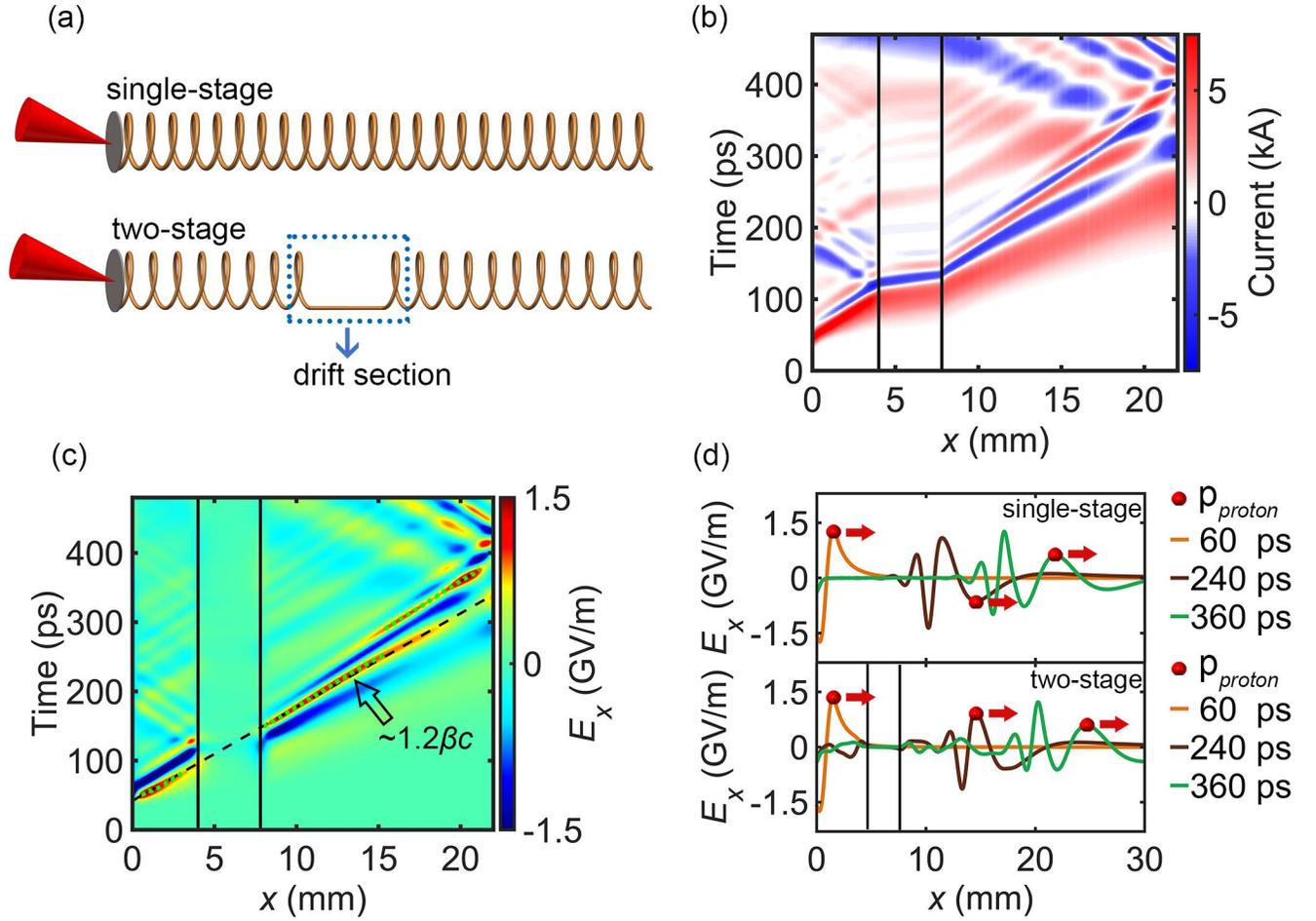

**Figure 6.** (a) The structure of a single-stage HC and a two-stage HC. (b) The spatial-temporal distribution of the current in the two-stage HC. (c) Longitudinal electric field in the two-stage HC, the black dashed line indicates the velocity mark of 1.2 $\beta c$, and the green dashed line marks extreme points of positive fields. (d) The $E_x$ and the position of protons with an initial energy of 25 MeV in single-stage (top layer) and two-stage (bottom layer), at 60 ps, 240 ps, and 360 ps, respectively. The red balls represent the positions of protons, and the vertical black lines in (b), (c), and (d) indicate the position of the drift section.

Figure 7(a) shows the protons' distribution in phase space and $E_x$ in the two-stage HC. The high-energy protons are kept in sync with the positive electric field at 60 ps, 240 ps, and 360 ps. As a result, their cut-off energies are significantly enhanced. Figure 7(b) shows the witnessed $E_x$ of the most energetic protons in their coordinate frame. They are accelerated in the first HC, and the cut-off energy is enhanced to 29.1 MeV at the position of $x = 4$ mm, slightly lower than that of the single-stage HC. Later, in the drift section, the $E_x$ is weak, and the cut-off energy does not change significantly. In the second HC, they are accelerated over a distance exceeding 20 mm in the emerged positive field. As the result of the overall 25 mm acceleration distance, the cut-off energy of the proton beam is increased to 45.1 MeV as compared to the final energy of 28 MeV in Figure 4(c).

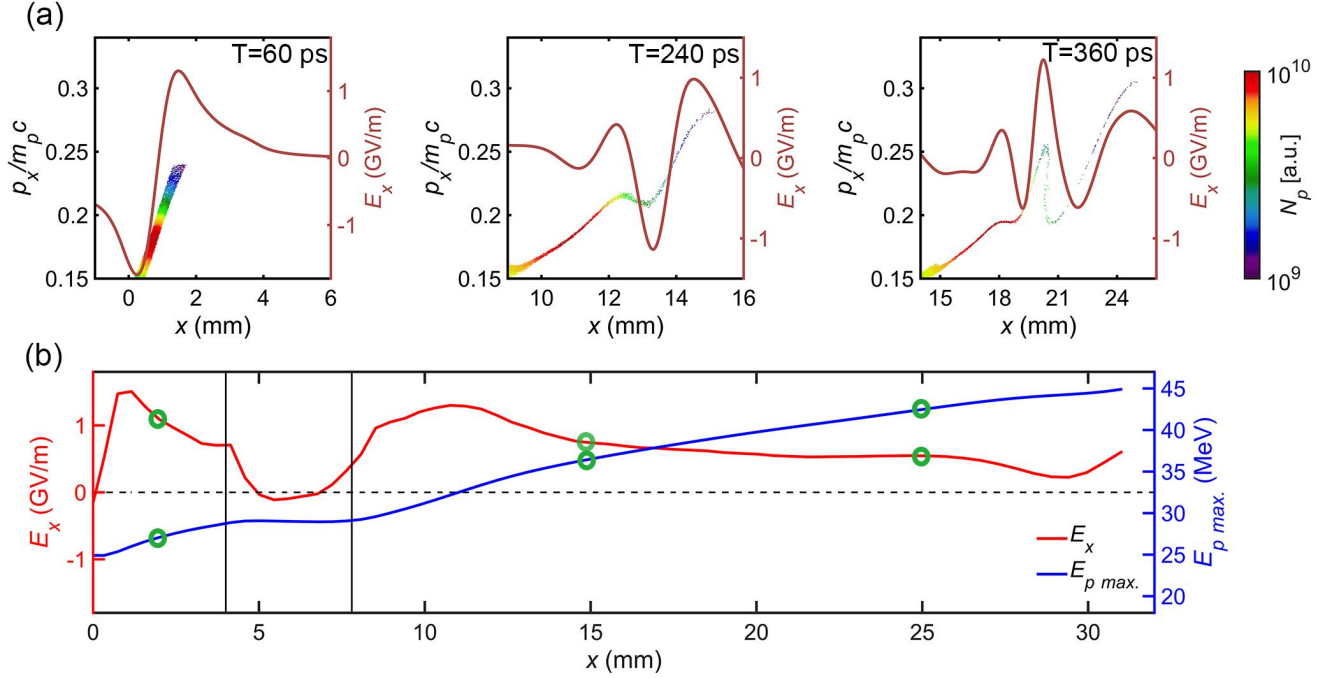

**Figure 7.** (a) Snapshots of the protons' phase space $(x, p_x)$ and $E_x$ at 60 ps, 240 ps and 360 ps, in the two-stage HC. (b) Longitudinal electric field in the coordinate frame of the traveling highest-energy protons (red curve) and the evolution of the maximum energy (blue curve) in two-stage HC, the three groups of green circles mark the three snapshots in (a).

The position and length of the drift section need to be delicately designed so that it can compensate for the delay of the electric field due to the phase reversal. We illustrate snapshots of the $E_x$ of the single-stage HC at different times and mark the spatial positions of protons in Figure 8(a). At 96 ps, when the most energetic protons are at $x = 4.2$ mm, the first and second positive electric fields are exactly equal. Thereafter, the amplitude of the first positive peak will be surpassed by the second one. This is a good point to let the protons be caught by the second positive field by introducing a drift section. For the sake of an integer number of coils turns, we introduce the drift section at $x = 4.0$ mm.

We also vary the length of the single-stage HC and that of the drift section in the two-stage HC, to investigate their energy gains, where the total length of the two-stage HC is kept at 32 mm. As shown in Figure 8(b), The maximum energy gain of the single-stage acceleration is 4.7 MeV with an HC length of 8 mm, which is already longer than the actual acceleration distance of the protons. The maximum energy gain of the two-stage HC achieves 20.1 MeV when the drift length is 3.8 mm. Moreover, energy gain over 15 MeV can be reached within ±1 mm deviation from the optimum drift length, making it robust for experiments. Figure 8(c) presents the simulated spectrum of the protons post-accelerated by the single-stage and two-stage HC. The primary protons are exponentially distributed with cut-off energy of 25 MeV. It is increased to 29.7 MeV with a single-stage HC of 8 mm length, and the cut-off energy is enhanced by 19%. Further increasing the length of the single-stage HC to 20 mm does not lead to an increase in the energy but a reduction. By using the two-stage HC scheme, the acceleration distance can be substantially increased. With the two-stage length of 32 mm, the proton cut-off energy can be enhanced to 45.1 MeV, an increase of 80.4% in cut-off energy, which is four times that of a single-stage HC.

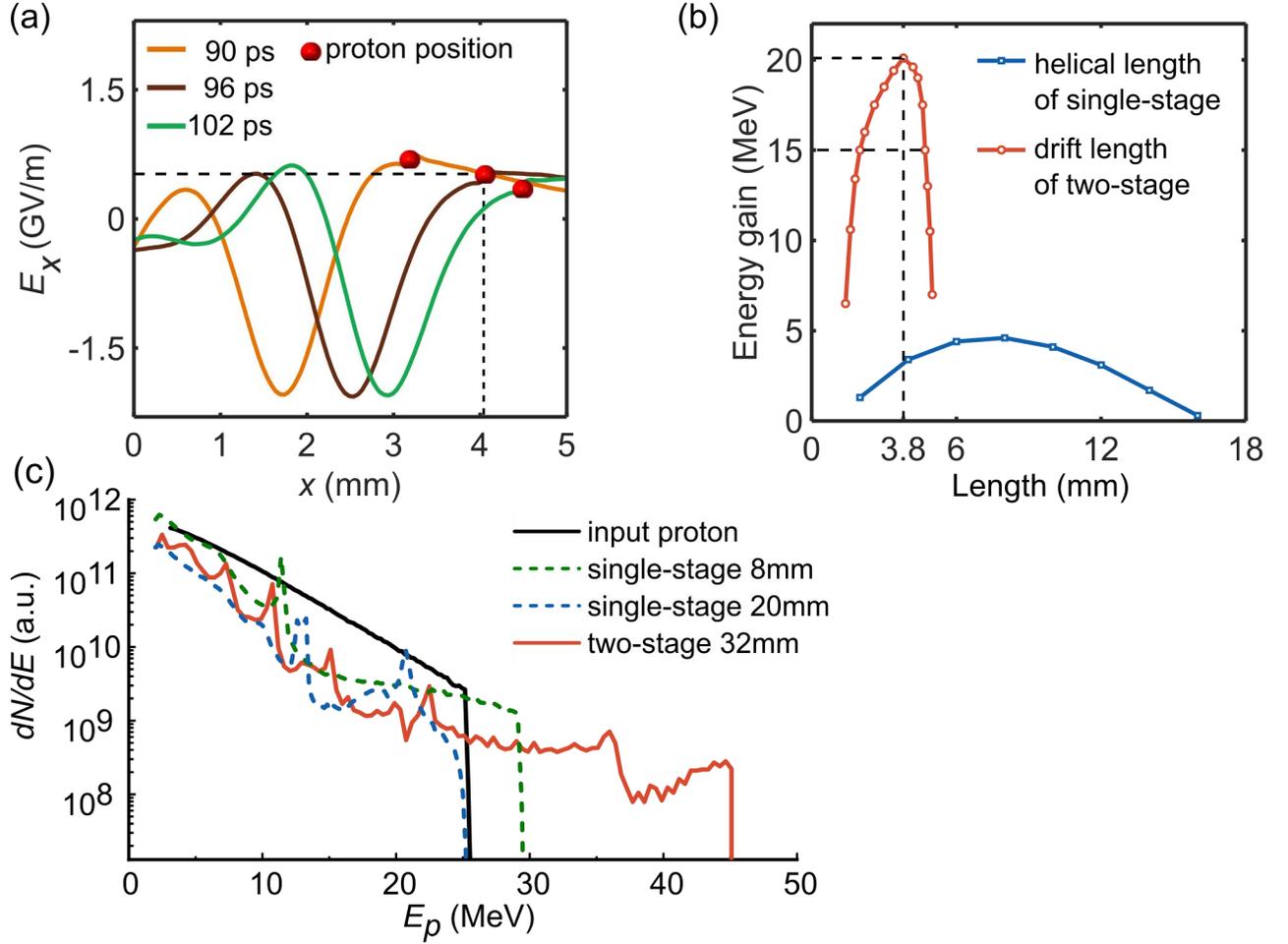

**Figure 8.** (a) snapshots of the current distribution and the positions of protons in HC, at 90 ps, 96 ps and 102 ps. (b) Energy gain by varying the helical length of single-stage and the drift length of two-stage HC. (c) Spectrum of the input protons (black line); spectrum after a single-stage HC of 8 mm (green dashed line) and 20 mm (blue dashed line); spectrum after a two-stage HC (red line).

We envision that the two-stage HC scheme can be utilized for a petawatt-class femtosecond laser to generate high-energy protons for the application of radiotherapy. As shown in Figure 9(a), the scaling for the charge of escaped electrons as a function of incident laser intensity can be obtained from the model reported by Poye *et al.*[40]. The scaling for the cut-off energy of protons is derived from the model of Dover *et al.*[41]. The red circles mark the hundreds-terawatt laser that we have simulated in the above works. We adopt a laser pulse width of 40 fs, a focal spot radius of 6 μm, and an absorption coefficient of 40% in the models. Hopefully, a total charge of escaped electron of 300 nC and a cut-off energy of protons of 60 MeV will be generated by a petawatt-class laser with intensity $\sim 3\times10^{21}$ Wcm$^{-2}$ (see red rhombuses in Figure 8(a)). A maximum longitudinal electric field on the central axis of HC of 3 GV/m could be formed. Protons with an exponential spectrum of protons with 60 MeV cut-off are injected into the HCs. According to the simulation results (Figure 9(b)), one would expect a cut-off of 72 MeV protons at the output of a single-stage HC. By employing a two-stage HC scheme, the cut-off proton energy can be enhanced to 102 MeV, which is sufficient to treat some shallow-seated tumors and most childhood cancers[46].

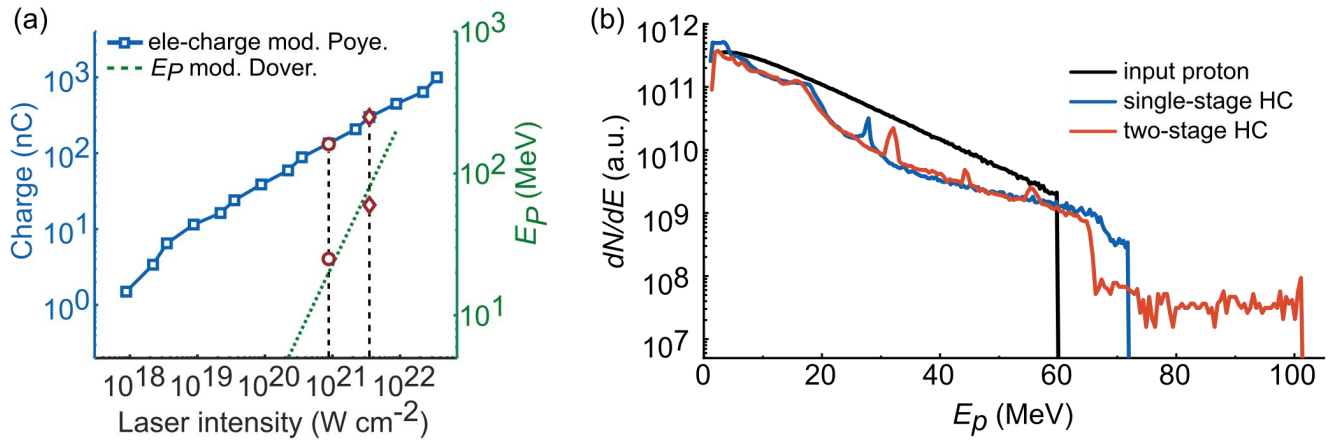

**Figure 9.** (a) Expected target charge of escaped electrons in the logarithmic scale calculated from the model in function of laser intensity (blue solid line). The green dashed line shows the cut-off energy of laser-driven protons against laser intensity. The requirements of the hundreds-terawatt laser and the petawatt laser in the simulations are marked out with red circles and rhombuses respectively. (b) Spectrum of the input protons in the simulations with the petawatt laser (black line); spectrum after a single-stage HC (blue line); spectrum after a two-stage HC (red line). The length of single-stage HC and two-stage HC are 10 mm and 40 mm, with $p = 0.55$ mm and $a = 0.3$ mm, and the drift length of 6.6 mm.

## 6. Discussion

The above results unambiguously demonstrate that a two-stage HC with a drift section is very beneficial for the enhancement of the proton energy. An interesting question would be: given that current dispersion and phase change could be controlled by the drift section, can a multi-stage scheme (over two stages) be applied to achieve longer distance synchronization and thus continue to boost proton energy?

We found that the energy enhancement in a multi-stage HC would be constrained due to the impedance mismatch between the coils and the drift sections. The impedance mismatch will cause a current reflection at the connection point between the coil and the drift section, as shown in Figure 10(a). The reflected current will superimpose on the original one to form a pulse train called reflection ringing (RR)[47]. We first conduct simulations of the current and the electric field in a structure consisting of three segments of HC based on the hundreds-terawatt laser. The current will reflect when it enters the drift section from the coil, resulting in disruption and weakening of the current (see Figure 10(b)). And in the third HC, the current distribution will become much more complicated due to the development of dispersion and RR.

The longitudinal electric field is also affected. We build the three-stage HC based on the two-stage HC according to the previously described strategy. The second drift section is introduced when the third positive electric field equals the second one. The delay between each positive electric field could be compensated as they are coincident with a black dashed line in Figure 10(c). So the synchronous acceleration of protons is maintained. However, the waveform of the electric field becomes more turbulent, and its strength gradually declines simultaneously. We check the energy gain from single-stage to four-stage HC separately in Figure 10(d). The blue curve shows the maximum strength of $E_x$ in the different stages. It is observed that it rapidly decreases as the number of stages increases, merely 0.5 GV/m at the fourth HC. As a result, the energy gain can be significantly raised to 20.1 MeV in the second HC but only modestly increases to 22 MeV or 23.2 MeV in the third or fourth HC, even though the total length of the HC has been doubled.

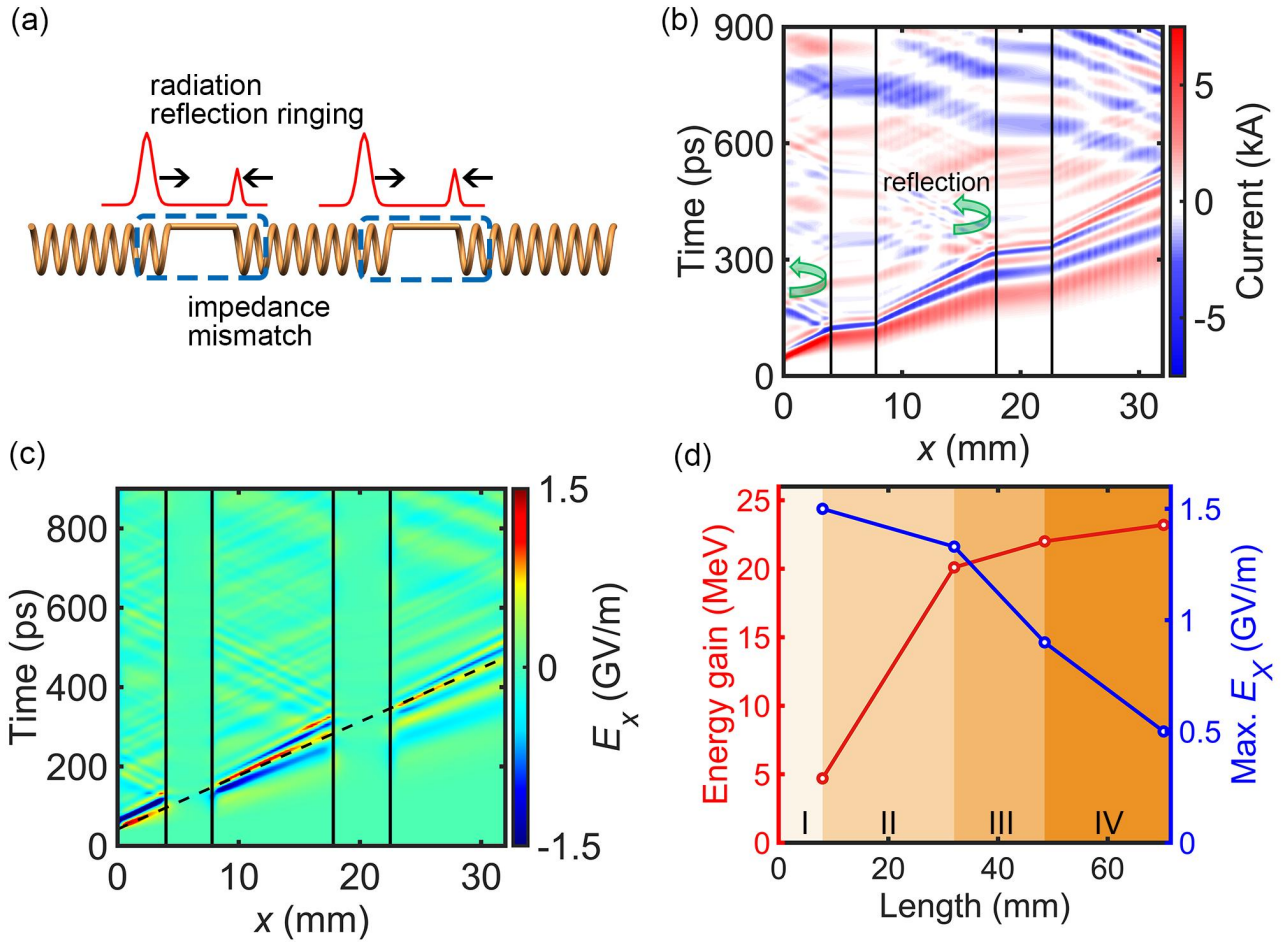

**Figure 10.** (a) Scheme of the reflection ringing of a three-stage structure HC. (b) Spatial-temporal distribution of the current in three-stage HC. (c) The spatial-temporal distribution of $E_x$ in three-stage HC, the black dashed lines in (c) mark the velocity of 1.2 $\beta c$, and the vertical black lines in (b) and (c) mean the drift sections. (d) Energy gain (red curves) and maximum intensity of $E_x$ (blue curves) for different stages of the HC (from 1 to 4 stages in different regions).

## 7. Conclusion

We demonstrate a dispersion-controlled scheme to enhance the energy of the post-acceleration protons by using a two-stage HC structure. The cut-off energy is improved from 25 MeV to 45.1 MeV with a hundreds-terawatt laser, four times higher than the single-stage HC. Over 100 MeV protons can be obtained by using a petawatt laser. Based on the self-consistent simulations and the circuit transmission model, for the first time, we reveal in detail how the transient current pulses disperse in HC, causing phase sliding and reversal in the electric field, and how the protons become desynchronous with the acceleration field. With a two-stage HC structure, the current transports faster in the drift section and can compensate for the dispersion-induced delay, thus enabling the proton and accelerated field to get synchronous again. In a conclusion, the two-stage scheme is a simple and practicable way to control the dispersion of the HCPA to enhance the energy gain of laser-driven ions, promising for the application of oncological therapy[48].

## Acknowledgement

This work was supported by the following projects: NSFC Innovation Group Project (grant number 11921006) and National Grand Instrument Project (grant number 2019YFF01014402). W. Ma acknowledges support from the National Science Fund for Distinguished Young Scholars (12225501).